\documentclass[10pt,twocolumn]{article}
\usepackage{epsf}

\title{
\large
\normalsize
\begin{center}
{\bf Thin Reaction Zones in Highly Turbulent Medium}
\end{center}
\normalsize
\small
\begin{center}
Vladimir A. Sabelnikov$^{1,2}$, Rixin Yu$^3$, and Andrei N. Lipatnikov$^4$
\end{center}
\vspace{-6mm}
\normalsize
\small
\begin{center}
{\it $^1$ONERA - The French Aerospace Lab., F-91761 Palaiseau, France \\
$^2$Central Aerohydrodynamic Institute (TsAGI), 140180 Zhukovsky, Moscow Region, Russian Federation \\
$^3$Division of Fluid Mechanics, Department of Energy Sciences, Lund University, 22100 Lund, Sweden \\
$^4$Department of Mechanics and Maritime Sciences, Chalmers University of Technology, 41296 Gothenburg, Sweden}
\end{center}
\vspace{1mm}
\begin{center}
\parbox{132mm}{
\small
\footnotesize
In a highly turbulent medium characterized by a low Damk\"ohler number $Da$, reactions are commonly considered to occur in 
distributed zones broadened by small-scale turbulent eddies.
In the present communication, an alternative regime of propagation of reaction waves in a highly turbulent medium is introduced and 
studied theoretically and numerically.
More specifically, propagation of an infinitely thin reaction sheet in a turbulent medium is analyzed, with molecular mixing of the
reactant and product being allowed in wide layers.
In this limiting case, an increase in the consumption velocity by turbulence is solely controlled by an increase in
the reaction-sheet area.
Based on physical reasoning and estimates, the area is hypothesized to be close to the mean area of an inert iso-scalar surface at
the same turbulent Reynolds number.
This hypothesis leads to a relation for the turbulent consumption velocity, which is similar to the well-known Damk\"ohler scaling
associated commonly with distributed reaction zones at a low $Da$.
The obtained theoretical results are validated by analyzing a big database (23 cases characterized by $0.01 \le Da <1$)
created recently in 3D direct numerical simulations of propagation of a statistically planar, one-dimensional, dynamically passive
reaction wave in statistically stationary, homogeneous, isotropic turbulence.
The DNS data well support the aforementioned relation.
They also show that the reaction is localized to thin zones even at $Da$ as low as 0.01, 
with a ratio of the turbulent and laminar consumption velocities being mainly controlled by the reaction-zone-surface area.
}
\end{center}
\date{}
\vspace{-10mm}
}

\begin{document}

\twocolumn
\maketitle
\vspace{-8mm}
\thispagestyle{empty}

\small
\begin{center}
{\bf I. INTRODUCTION}
\end{center}

A problem of the influence of turbulence on reaction waves is straightforwardly relevant to various
phenomena ranging from reactions in aqueous solution \cite{RHR95}, combustion [2-6],
and deflagration-to-detonation transition \cite{PGO11,CAL13} under terrestrial conditions to evolution of thermonuclear Ia supernovae
\cite{GKOCR03,GKO04} in the Universe.
This non-linear and multi-scale problem attracted much attention since 1940s when significant acceleration of flame propagation
by turbulence was found.
The effect was first explained by Damk\"ohler \cite{Da} who hypothesized two limiting regimes of the influence of turbulence on a
flame depending on a ratio of an integral turbulence length scale $L$ to the laminar flame thickness $\delta_L$.
Subsequently, various regimes of the influence of turbulence on a reaction wave were widely discussed, with a ratio of
the rms turbulent velocity $u'$ to the laminar wave speed $S_L$ being considered to be another (in addition to $L/\delta_L$) 
important number for identifying such regimes.
For instance, several regime diagrams of premixed turbulent combustion were invented [12-18]
and are widely used now.

While various regimes of the influence of turbulence on a chemical reaction are well studied, there is an important
exception, i.e., a reaction wave at a low Damk\"ohler number $Da=\tau_T/\tau_L \ll 1$, which characterizes a ratio of
the turbulence and laminar-reaction-wave time scales, $\tau_T=L/u'$ and $\tau_L=\delta_L/S_L$, respectively.
On the one hand, instantaneous reaction zones are commonly assumed to be
``distributed" \cite{FAW}, ``thickened" \cite{Bor84,VV02}, ``broadened" \cite{Dr08}, or ``well-stirred" \cite{Pet21} 
at $Da \ll 1$.
The concept of distributed reactions is often considered to be validated by the fact that the same square-root dependence of the
normalized turbulent consumption velocity $U_T/S_L$ on the turbulent Reynolds number $Re_t=u' L/\nu$ was predicted by Damk\"ohler 
\cite{Da} in
the case of $L \ll \delta_L$, associated widely with the distributed reactions, and was documented in experimental studies of
turbulent reacting flows \cite{RHR95,SLLCLY35}, as well as in DNS's of thermonuclear turbulent combustion \cite{ABW10}.
Here, $U_T$ is the mean rate of consumption of reactants per unit area of a mean reaction-wave surface and
$\nu$ is the kinematic viscosity.

On the other hand, there are experimental and DNS data that put the concept of distributed reactions 
or, at least, the boundary of such a regime, into question.
Indeed, by reviewing experiments with premixed flames, Driscoll \cite{Dr08} did not find a laser-diagnostic result that showed
significantly broadened heat-release zones.
In the latest experimental [21-24]
or DNS [25-28]
studies of flames characterized by a low $Da$, significantly broadened heat-release zones were not detected either.

Accordingly, the present communication aims at clarifying the issue by
(i) introducing another regime of the influence of turbulence on a reaction wave at $Da \ll 1$ and
(ii) revealing physical mechanisms that control an increase in $U_T/S_L$ by $Re_t$ in this regime.
Other goals are
(iii) to derive the well-known analytical relation for $U_T \propto S_L Re_t^{1/2}$ by considering the newly introduced regime and, 
therefore, 
(iv) to reconcile the experimental \cite{RHR95,SLLCLY35} and DNS \cite{ABW10} data that support this relation with 
the experimental [21-24]
and DNS [25-28]
data that show thin heat release zones in 
turbulent flames characterized by a low $Da$. 
Yet another goal is 
(v) to validate the theoretical study by recent DNS data 
[29-31].
New results obtained by analyzing the data will be discussed first (Sect. II) to show a need for the subsequent theoretical analysis 
(Sect. III).

It is worth noting that the present study addresses a constant-density reaction wave. 
Nevertheless, the following analysis and results are also relevant to combustion regime diagrams and distributed burning, 
because both the most recognized diagrams [13-15]
and the paradigm of distributed reactions were developed by neglecting density variations in flames [11,13-15].

\begin{center}
{\bf II. DIRECT NUMERICAL SIMULATIONS \label{SDNS}}
\end{center}

We will restrict ourselves to a brief summary of the DNS attributes and techniques, because they were already discussed in detail in 
other papers [29-31]
that the interested reader is referred to.
Propagation of a statistically planar, 1D, constant-density, single-reaction wave in forced, statistically stationary,
homogeneous, isotropic turbulence was numerically investigated by solving the 3D mass-conservation ($\nabla \cdot \mathbf{u}=0$)
and Navier-Stokes equations, as well as the following transport equation
\begin{eqnarray}
\label{EPVBE}
\frac{\partial c}{\partial t} + \mathbf{u} \cdot \nabla c = D \Delta c + W 
\end{eqnarray}
for a reaction progress variable $c$, i.e., mass fraction of a reactant or product species normalized so that $c=0$ and 1
in fresh reactants and equilibrium products, respectively.
Here, $t$ is time, $\mathbf{u}$ is the flow velocity vector, $D$ is the molecular diffusivity of $c$,
\begin{eqnarray}
\label{EW}
W = \frac{1}{1+\tau} \frac{1-c}{\tau_R} \exp \left[- \frac{Ze (1+\tau)^2}{\tau (1+\tau c)} \right]
\end{eqnarray}
is the reaction rate, $\tau=6$, and
two values of the Zel'dovich number $Ze=6.0$ and 17.1 were used to change the thickness $\delta_r$ of the reaction,
with $S_L$ and $\delta_L=D/S_L$ retaining their values due to appropriate adjustment of the time scale $\tau_R$.
The larger $Ze$, the more non-linear the dependence of $W$ on $c$ and the less $\delta_r$.
Here, $\delta_r=(x_2-x_1)/(c_2-c_1)$, the boundaries $c_1$ and $c_2$ of the reaction zone are set by
$W(c_1)=W(c_2)=0.5 \max{\{W(c)\}}$, and $x_2-x_1$ is the spatial distance between these boundaries in the planar, 1D
laminar reaction wave.

The boundary conditions were fully periodic.
The thickness of the entire reaction wave brush was always significantly less than the length $\Lambda_x$ of the computational domain.
Statistics were sampled over a time interval of $\theta>50 \tau_T^0$ after the end $t_0 > 3\tau_T^0$ of a transient phase.
Mean value $\overline{q}$ of a quantity $q$ was averaged over transverse planes ($x$=const) and time at $t_0 < t \le t_0+\theta$.

Various cases were set up by specifying $S_L$, $\delta_L$, $Ze$, and the width $\Lambda$ of a rectangular computational domain,
with $\Lambda$ controlling $L$ and $Re_t$.
The values of $D$ and $\tau_R$, required to obtain the specified $S_L$ and $\delta_L$, were found in 1D pre-computations of
the laminar wave.
Since the reaction wave did not affect the flow, the choice of $\Lambda$ (and, hence, $L$ and $Re_t$) was independent of
the choice of $S_L$, $\delta_L$, and $Ze$.
Totally 45 cases characterized by $Da=0.01-24.7$,
$u'/S_L=0.5-90$, and $L/\delta_L=0.39-12.4$ were simulated,
with a few cases being designed to show weak sensitivity of computed results to grid resolution, $L/\Lambda$, etc.
16 non-dimensional characteristics of each of the 45 cases are reported in Table I in Ref. \cite{YLPoF17}, where
reasons for selecting those cases  are discussed in detail.
In the following, we will solely consider 23 cases characterized by $Da<1$.
In these 23 cases, the Karlovitz number
\begin{equation}
Ka = \frac{\tau_L}{\tau_K} = Sc \left( \frac{\delta_L}{\eta_K} \right)^2 = \frac{1}{Sc} \left( \frac{v_K}{S_L} \right)^2 
= \frac{Re_t^{1/2}}{Da} 
\end{equation}
was varied from 6.5 to 587.
Here, the Kolmogorov time $\tau_K=(\nu/\overline{\varepsilon})^{1/2}$, length $\eta_K=(\nu^3/\overline{\varepsilon})^{1/4}$, and
velocity $v_K= \eta_K/\tau_K$ scales characterize the smallest turbulent eddies,
$\overline{\varepsilon} \propto {u'}^3/L$ is the mean dissipation rate \cite{MY,LL}, and
$Sc=\nu/D$ is the Schmidt number.

\begin{figure} [t]
\begin{center}
\leavevmode \epsfysize=50.0 mm
\epsfbox{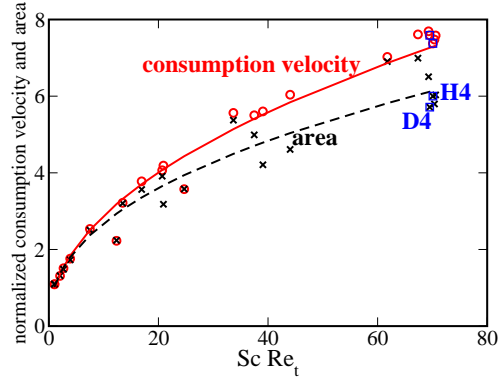}
\end{center}
\vspace{2mm}
\caption{\small
Normalized turbulent consumption velocity $U_T/S_L$ (red circles) and reaction-surface area $\langle \delta A \rangle$ (black crosses)
vs. $Sc Re_t$.
Red solid and black dashed lines fit the DNS data with $U_T/S_L=0.96 (Sc Re_t)^{0.48}$ and
$\langle \delta A \rangle = (Sc Re_t)^{0.43}$, respectively.
Blue squares mark DNS data obtained in cases D4 and H4 characterized by $Ze=6.0$ and 17.1, respectively,
with all other things ($Da, \ Ka, \ u'/S_L, \ Sc$, etc.) being equal.
}
\label{FUt}
\end{figure}

Computed turbulent consumption velocities
\begin{eqnarray}
\label{EUtDNS}
U_T = \int_{0}^{\Lambda_x} \overline{W} \mathrm{d} x
\end{eqnarray}
are shown in circles in Fig. \ref{FUt}, with solid curve fitting the DNS data. 
The data agree very well with the classical Damk\"ohler's scaling \cite{Da} of $U_T \propto S_L Re_t^{1/2}$.
Crosses and fitting dashed curve show a relative increase in the mean area of the reaction-zone surface, evaluated as follows
\begin{eqnarray}
\label{EdA}
\langle \delta A \rangle = \frac{1}{\theta \Lambda^2} \int_{t_0}^{t_0+\theta} \int_0^{\Lambda_x} \int_0^{\Lambda} \int_0^{\Lambda}
\frac{|\nabla c| \Pi(c)}{c_2-c_1} \mathrm{d} \mathrm{x} \mathrm{d} t,
\end{eqnarray}
where $\Pi(c) = \mathrm{H}(c-c_1)-\mathrm{H}(c-c_2)$ is difference between Heaviside functions.
Crosses and circles in Fig. \ref{FUt} indicate that values of $\langle \delta A \rangle$ and $U_T/S_L$ are
comparable in all cases, contrary to the paradigm of distributed reactions, which attributes an increase in $U_T/S_L$ to 
turbulent mixing within broadened reaction zones, rather than an increase in the area $\langle \delta A \rangle$.

\begin{figure} [t]
\begin{center}
\leavevmode \epsfysize=50.0 mm
\epsfbox{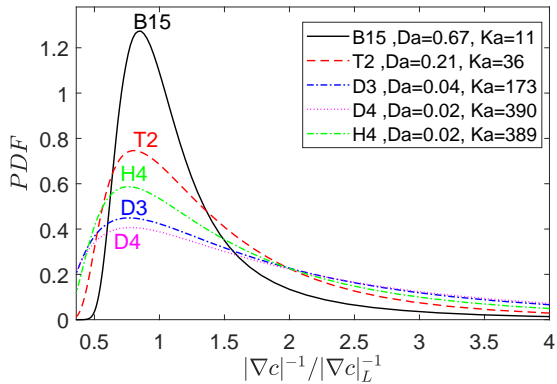}
\end{center}
\vspace{2mm}
\caption{\small
Probability density functions for the normalized local reaction-zone thickness $|\nabla c|_L/|\nabla c|_r$
computed in five cases specified in legends.
Here, $|\nabla c|_r$ is conditioned to $c_w-0.005 < c < c_w+0.005$
and $c_w$ is associated with the peak $W(c)$.
The laminar $|\nabla c|_L$ is evaluated at $c=c_w$.
$Ze=17.1$ in case H4 and $Ze=6.0$ in other cases.
}
\label{FPDF}
\end{figure}

Figure \ref{FPDF} also puts the paradigm into question by showing that probability density functions for the normalized local
reaction-zone thickness $|\nabla c|_L/|\nabla c|_r$ peak in the vicinity of unity at various $Da$, including $Da \ll 1$.
While probability of finding $|\nabla c|_L/|\nabla c|_r > 2$ increases with decreasing $Da$, this probability is low,
i.e., broadening of reaction zones is weakly pronounced.
An increase in $Ze$ reduces the probability, see cases D4 and H4.
Moreover, there is a substantial probability of finding locally thinned reaction zones characterized by
$|\nabla c|_L/|\nabla c|_r<1$.

Thus, the present DNS data are consistent not only with
(i) the classical Damk\"ohler expression of $U_T \propto S_L Re_t^{1/2}$ and
(ii) experimental \cite{RHR95,SLLCLY35} and DNS \cite{ABW10} data that support it, but also with
(iii) the latest experimental [21-24]
and DNS [25-28]
data that show thin heat release zones in turbulent flames characterized by a low $Da$.
Accordingly, the cited papers 
and the present DNS data call for development of an alternative (to the paradigm of distributed reactions) concept of the influence
of intense turbulence on a reaction wave at low Damk\"ohler numbers.
This call will be responded in the next section.

Finally, it is worth noting that the obtained values of $U_T/S_L$ (or $\langle \delta A \rangle$) are almost the same in
cases D4 and H4, see blue squares in Fig. \ref{FUt}.
Because the sole difference between the two cases consists in different reaction-zone thicknesses ($Ze=6.0$ and 17.1, respectively),
whereas the values of $Da$ (or $Ka$, $u'/S_L$, $Sc$, etc.) are equal in both cases,
these DNS data indicate that neither $U_T/S_L$ nor $\langle \delta A \rangle$ depends on the reaction zone thickness at a very low
$Da=0.02$.
This apparently surprising result will be explained in the next section.

\begin{center}
{\bf III. THEORETICAL STUDY \label{ST}}
\end{center}

Similar to the DNS, 
let us consider a statistically planar, 1D reaction wave that propagates from right to left along $x$-axis
in 3D homogeneous isotropic turbulence, but does not affect it.
We will address the case of a low $Da \ll 1$, a high $Re_t \gg 1$, and, consequently, a high $Ka=Re_t^{1/2}/Da \gg 1$.

To explore a physical scenario opposite to the widely accepted paradigm of distributed reactions, let us study a reaction whose rate
$W$ depends on $c$ in the extremely non-linear manner, i.e., $W[c(\mathbf{x},t)]$ is localized to an infinitely thin reaction sheet,
but vanishes outside it.
Such a limiting case is often studied, e.g., in the combustion literature following 
the pioneering ideas by Zel'dovich and Frank-Kamenetskii who developed the well-known
ZFK theory of laminar premixed flames \cite{ZBLM} by analyzing an asymptotic solution to Eq. (\ref{EPVBE}) at large $Ze$ in 
Eq. (\ref{EW}). 

In the considered case, the following constraints hold at the reaction sheet, e.g., \cite{ZBLM,Kli63},
\begin{eqnarray}
\label{EBC}
c|_r=1, 
\hspace{5mm} 
| \mathbf{n} \cdot \nabla c|_r = |\nabla c|_r = \left| \frac{\partial c}{\partial n} \right|_r 
= \frac{S_L}{D} = \frac{1}{\delta_L},
\end{eqnarray}
i.e., the reaction progress variable is continuous, but its gradient drops from $\delta_L^{-1}$ on the reactant side of the reaction 
sheet to zero on the product side.
Under the above assumptions, the structure of a reaction wave in a turbulent flow is modeled by a standard transport equation
\begin{eqnarray}
\label{EPVBEi}
\frac{\partial c}{\partial t} + \mathbf{u} \cdot \nabla c = D \Delta c 
\end{eqnarray}
provided that Eq. (\ref{EBC}) holds at the reaction sheet and the boundary condition of $c(-\infty,y,z,t)=0$
is set far ahead of the sheet.
Here, 
$\mathbf{n} = - \nabla c/|\nabla c|$ is the unit vector normal to an iso-surface $c(\mathbf{x},t)=$const and
$n$ is spatial distance counted from the reaction sheet along the $\mathbf{n}$-direction.

Equation (\ref{EBC}) warrants that the reactant flux $D | \partial c/\partial n |_r$ towards the reaction sheet is equal to
the rate $S_L$ of the reactant consumption per unit sheet area,
but the speed $(S_d)_r = D (\Delta c/|\nabla c|)_r$ of self-propagation of the sheet in an inhomogeneous
flow can significantly differ from $S_L$ \cite{Kli63}.
For example, if term $D \Delta c$ is rewritten in the spherical coordinate framework, the speed $(S_d)_r$ involves
an extra term whose magnitude $2 D/r_r$ can be much larger than $S_L$ if the curvature
radius $r_r$ of the sheet is small, i.e., $r_r \ll \delta_L$.
Strong variations in $(S_d)_r$ can also be caused by local velocity gradients 
\cite{Kli63}.

The problem stated above is fully consistent with the following two features of highly turbulent reaction waves, 
discussed in Sect. II.
First, the reaction is localized to thin zones in the simulations or to infinitely thin sheets within the framework of that problem.
Second, a ratio of $U_T/S_L$ is mainly controlled by $\langle \delta A \rangle$ in the simulations and is solely controlled by 
the area increase within the framework of the problem.
Accordingly, the specific goals of the following study are 
(i) to obtain the classical expression of $U_T \propto S_L Re_t^{1/2}$ and
(ii) to explain weak influence of $Ze$ on $U_T/S_L$ by studying the stated problem.

Before doing so, it is worth noting that 
this problem differs fundamentally from a problem of front propagation in a turbulent medium [36-38],
because molecular diffusion, i.e., a term on the Right Hand Side (RHS) of Eq. (\ref{EPVBEi}), is not directly addressed in the latter case.
Accordingly, the latter problem is associated with $L \gg \delta_L$ and $Da \gg 1$, whereas the present communication addresses
the case of a low $Da$.
It is worth stressing that molecular diffusion smooths out small-scale wrinkles of the reaction sheet, generated by turbulent
eddies, and, therefore, significantly reduces turbulent consumption velocity $U_T$ \cite{YLPRE17}
when compared to a linear dependence of $U_T \propto u'$ in the case of front propagation \cite{YBL15}.
For example, if $Sc=\mathrm{O}(1)$ and the smallest turbulent eddies wrinkle a reaction sheet so that the local curvature radius
$r_r=\mathrm{O}(\eta_K)$,
then, the diffusion contribution $\mathrm{O}(D/\eta_K)$ to the speed $(S_d)_r$ is comparable with
the magnitude $v_K$ of velocity fluctuations associated with such eddies and is much larger than $S_L$ at $Ka \gg 1$.
Since the eddy lifetime $\tau_K$ is short, but molecular diffusion affects the wrinkle even after dissipation of the eddy,
small-scale (when compared to $\delta_L$) wrinkles are efficiently smoothed out by molecular diffusion
\cite{YLPRE17}.

Within the framework of the problem stated above, the reaction term vanishes everywhere with exception of the reaction sheet and
evolution of an iso-surface $c(\mathbf{x},t)=C<1$ is described by the diffusion Eq. (\ref{EPVBEi}) with the boundary 
conditions set by Eq. (\ref{EBC}).
The concentration gradient at the reaction sheet $|\nabla c|_r = \delta_L^{-1}$ is solely determined by the reaction 
and molecular diffusion, but does not depend on turbulence.
When distance from the reaction sheet is increased, the turbulence begins affecting the concentration gradient, whereas 
the influence of the boundary condition is reduced.
Accordingly, it is reasonable to assume that, at some distance from the reaction sheet, the turbulence overwhelms the influence of 
the reaction and the local concentration gradients are solely controlled by inert turbulent advection and molecular transport.
At $Da \ll 1$, the thickness of such a transition layer attached to the  reaction sheet is expected to be much smaller than 
the Kolmogorov length scale.

This hypothesis can be supported by order-of-magnitude estimates discussed below.
It is worth stressing, however, that such estimates given by Eqs. (\ref{Egrcgrc})-(\ref{Enr}) are not more than just estimates 
that aim solely at supporting the hypothesis.
In other words, Eqs. (\ref{Egrcgrc})-(\ref{Enr}) are not claimed to be theoretical expressions.
The major theoretical result of the following analysis consists in Eq. (\ref{EUT}), which will be obtained with qualitative rigor 
based on the hypothesis that is justified in particular by the order-of-magnitude estimates given by Eqs. (\ref{Egrcgrc})-(\ref{Enr}).

First, let us compare the magnitude of the concentration gradient at the reaction sheet, i.e.,
$|\nabla c|_r = \delta_L^{-1}$, with the magnitude $|\nabla c|_T$ of the concentration gradients outside the transition layer.
The order of the latter magnitude may be estimated based on the widely accepted view that, in an inert flow,
the mean scalar dissipation rate $\overline{N}=D \overline{\left( \nabla c \right)^2}$
is independent of turbulent Reynolds number provided that $Re_t \gg 1$ [40-42],
i.e., $\overline{N} \propto 1/\tau_T$.
Accordingly, the magnitude of $|\nabla c|_T$ scales as
$(D \tau_T)^{-1/2} \propto (Sc Re_t)^{1/2}/L$
and a ratio of the magnitudes of the two gradients may be estimated as follows
\begin{eqnarray}
\label{Egrcgrc}
\frac{|\nabla c|_r}{|\nabla c|_T} \propto \frac{S_L}{D} L \left( Sc Re_t \right)^{-1/2} \propto Da^{1/2} \ll 1.
\end{eqnarray}
The scalar dissipation rate vanishes in the products, jumps to $D |\nabla c|_r^2 = \tau_L^{-1}$ at the reaction sheet, increases further in
the transition layer, and is on the order of $\tau_T^{-1} = (\tau_L Da)^{-1} \gg \tau_L^{-1}$ outside the layer.
Thus, if $Da \ll 1$ and $Re_t \gg 1$, turbulent mixing overwhelms the effect of the reaction on the
$c(\mathbf{x},t)$-field everywhere with exception of a narrow layer close to the reaction sheet.

Second, to estimate the thickness of the transition layer, let us expand $c(n)$ to Taylor series in the vicinity of
the reaction sheet, i.e.,
\begin{eqnarray}
\label{ETS}
c = 1 - \left| \frac{\partial c}{\partial n} \right|_r n
+ \frac{1}{2} \left. \frac{\partial^2 c}{\partial n^2} \right|_r n^2
+ \mathrm{O}(n^3).
\end{eqnarray}
Since the expansion coefficient in the linear term is solely controlled by the reaction, see Eq. (\ref{EBC}),
let us assume that the thickness $n_r$ of the considered layer may be estimated by equating the linear and quadratic terms in Eq. 
(\ref{ETS}).
To estimate $\left| \partial^2 c/\partial n^2 \right|_r$ in a turbulent flow, let us consider the simplest relevant model problem,
i.e., an 1D planar laminar reaction wave stabilized in a 2D flow $\{u=-\gamma x,v=\gamma y\}$, with the velocity gradient $\gamma$
being on the order of $\tau_K^{-1}$ \cite{Kli63,KuSa77}.
In such a case, Eq. (\ref{EPVBEi}) reads
\begin{eqnarray}
\label{EPVBE1D}
- \frac{x}{\tau_K} \frac{\mathrm{d} c}{\mathrm{d} x} = D \frac{\mathrm{d}^2 c}{\mathrm{d} x^2},
\end{eqnarray}
Eq. (\ref{EBC}) holds, $n=x_r-x$, and $c(-\infty)=0$.
Integration of Eq. (\ref{EPVBE1D}) results in $x_r \approx Sc^{-1/2} \eta_K \sqrt{\ln(Ka/2 \pi)}$ if $Ka \gg 1$,
i.e., if $\tau_K \ll \delta_L^2/D$ \cite{KuSa77}.
Consequently, Eqs. (\ref{EBC}), (\ref{ETS}), and (\ref{EPVBE1D}) yield
\begin{eqnarray}
\label{Enr}
n_r \approx 
\frac{2 {\eta_K}}{\sqrt{Sc \ln(Ka/2 \pi)}} \ll Sc^{-1/2} \eta_K.
\end{eqnarray}
At the boundary of the transition layer, the difference between unity and the boundary value $c^*$ of the reaction progress variable 
is less than $\epsilon = 2/\sqrt{Ka \ln(Ka/2 \pi)} \ll 1$, see the second term on the RHS of Eq. (\ref{ETS}) 
and note that the positive third term makes the difference even smaller.

The above order-of-magnitude estimates support the following scenario.
If $Da \ll 1$ and $Re_t \gg 1$, the reaction significantly affects the $c(\mathbf{x},t)$-field in a narrow layer 
($c^*<c<1$ with $c^* > 1 - \epsilon$) in the vicinity of the reaction sheet. 
The thickness of this layer is less than $\eta_K$ if $Sc=\mathrm{O}(1)$.
Since distance between the inert iso-surface $c(\mathbf{x},t)=c^*<1$ and the reaction sheet is so 
small, we may assume that the two surfaces move in a close correlation with one another and, hence,
their areas are roughly equal, i.e., $A_r \approx A_{c^*}$.
The latter area can be estimated as follows $A_{c^*} \propto A_c$ invoking knowledge on the area $A_c$ of an iso-scalar surface in
the case of inert turbulent mixing.
This is the key point of the present concept.

By analyzing experimental data on inert turbulent mixing, Kuznetsov and Sabelnikov \cite{KuSa} hypothesized independence of the
probability density function $P(c)$ and mean scalar dissipation rate 
on $Re_t \gg 1$ and arrived at
\begin{eqnarray}
\label{EAc}
A_c \propto A_0 Re_t^{1/2}
\end{eqnarray}
if $Sc=$const.
Here, $A_c$ and $A_0$ are areas of instantaneous, $c(\mathbf{x},t)=C$, and mean, $\bar{c}(\mathbf{x},t)=C$,
iso-surfaces, respectively.
The same scaling results from widely accepted independence \cite{SRM89,CHAKKC13} of the bulk inert scalar flux $F_c$ through an
iso-surface $c(\mathbf{x},t)=$const on $Re_t \gg 1$.
Indeed, since
\begin{eqnarray}
\label{EFc}
F_c = D \int_{A_c} \left| \frac{\partial c}{\partial n} \right|_T \mathrm{d} A_c 
\propto \frac{D}{L} \left( Sc Re_t \right)^{1/2} A_c 
\nonumber \\
\propto u' A_c \left( Sc Re_t \right)^{-1/2}, 
\end{eqnarray}
$A_c$ should be proportional to $(Sc Re_t)^{1/2}$
in order for the flux $F_c$ to be independent of $Re_t$.
It is worth noting that the flux $F_c$ is controlled by the relative velocity of the surface $c(\mathbf{x},t)=$const with
respect to the local flow, with the relative velocity being solely controlled by the molecular diffusion.
Accordingly, in the case of a material surface, $D=F_c=0$ and Eqs. (\ref{EAc}) and (\ref{EFc}) do not hold.

Finally, Eq. (\ref{EBC}) and $A_c/A_0 \propto (Sc Re_t)^{1/2}$ yield
\begin{eqnarray}
\label{EUT}
U_T = 
S_L \frac{A_r}{A_0} \propto S_L \frac{A_c}{A_0} \propto S_L \left( Sc Re_t \right)^{1/2}  
\end{eqnarray}
for a turbulent consumption velocity $U_T$, i.e., bulk consumption rate divided with the area $A_0$ of the mean reaction-wave
surface.

Equation (\ref{EUT}) is well supported by the DNS data plotted in Fig. \ref{FUt}, 
is consistent with experimental \cite{RHR95,SLLCLY35} and DNS \cite{ABW10} data on $U_T$,
and is also consistent with experimental [18,21-24]
and DNS [25-28]
observations of thin reaction zones at a low $Da$.
While Eq. (\ref{EUT}) coincides with the classical scaling \cite{Da}, the two results
were obtained (i) invoking different assumptions, i.e., $L \ll \delta_L$ \cite{Da} and $Da \ll 1$ in the present work, and
(ii) for different regimes of turbulent wave propagation, i.e., distributed reactions \cite{Da}, but infinitely thin reaction zones 
in the present work.
The governing physical mechanisms of the influence of turbulent eddies on $U_T$ are different in the two cases, 
i.e., intensification of mixing within broad reaction zones and an increase in the area of the reaction sheet, respectively.
A common feature of the two approaches consists in highlighting turbulent mixing.

It is also worth stressing that the present analysis is fully consistent with DNS data that are commonly considered to prove the 
concept of distributed reactions by showing significant difference between Joint Probability Density Functions (JPDF's) for fuel and 
temperature, obtained from thermonuclear laminar and highly turbulent flames \cite[Fig. 1]{ABW10}.
The point is that, within the framework of the present concept, the reaction can affect such a JPDF solely within a narrow transition
layer. Outside the layer, the JPDF is controlled by turbulent mixing, in line with the discussed DNS data. 
Moreover, in line with the present concept, the DNS data do show that the turbulent and laminar JPDF's are close to one another 
within a narrow layer adjacent to the product side of the simulated thermonuclear flames \cite[Fig. 1]{ABW10}.

Since the developed concept uses the area of an inert iso-scalar surface to estimate the reaction-sheet area, 
the two areas should be independent of the reaction-zone structure in the case of a thin reaction zone of a finite thickness,
thus, explaining the DNS data that show a very weak (if any) dependence of $\langle \delta A \rangle$ on the reaction-zone thickness 
at $Da=0.02$, see blue squares in Fig. \ref{FPDF}.
%
%
\vspace{33mm}

\begin{center}
{\bf IV. CONCLUSIONS \label{SC}}
\end{center}

Propagation of an infinitely thin reaction sheet in a turbulent medium is analyzed in the case of $Da \ll 1$.
By allowing for molecular mixing, a relation for the turbulent consumption velocity $U_T$, see Eq. (\ref{EUT}), is obtained.
Within the framework of the analysis, an increase in the consumption velocity by turbulence is controlled by an increase in
the reaction-sheet area, which is argued to be close to the mean area of an inert iso-scalar surface at the same $Re_t$.
DNS data obtained in 23 cases characterized by $0.01 \le Da < 1$ validate the derived Eq. (\ref{EUT})
and confirm the aforementioned governing physical mechanism of the increase in $U_T$ by turbulence.
Moreover, the DNS data show that, even at a low $Da$, the reaction zone thickness is statistically close to the thickness of the
reaction zone in a laminar flow.
Furthermore, the DNS data indicate that $U_T$ is weakly affected by the reaction-zone thickness,
thus, further validating the theory.

The present analysis and DNS data show that, even at a low $Da$, 
reaction zone may be thin and an increase in the consumption velocity by turbulence may be controlled by an increase in the 
area of the reaction surface, with the latter increase being well described by the theory of inert turbulent mixing.  
The obtained results offer an opportunity to reconcile experimental \cite{RHR95,SLLCLY35} and DNS \cite{ABW10} data that support 
the classical Damk\"ohler expression, i.e., $U_T/S_L \propto Re_t^{1/2}$, with the latest 
experimental [21-24]
and DNS [25-28]
data that show thin heat release zones in turbulent flames characterized by a low $Da$.
\vspace{3mm}

\begin{center}
{\bf ACKNOWLEDGMENTS}
\end{center}
\vspace{3mm}

VAS gratefully acknowledges the financial support by ONERA and by the Grant of the Ministry of Education and
Science of the Russian Federation (Contract No. 14.G39.31.0001 of 13.02.2017).
RY gratefully acknowledges the financial support by the Swedish Research Council.
ANL gratefully acknowledges the financial support by the Chalmers Transport and Energy Areas of Advance, and
by the Combustion Engine Research Center.

\end{document}